\documentclass[10pt,a4paper,english]{article}\usepackage[]{graphicx}\usepackage[]{color}
\makeatletter
\def\maxwidth{ %
  \ifdim\Gin@nat@width>\linewidth
    \linewidth
  \else
    \Gin@nat@width
  \fi
}
\makeatother

\definecolor{fgcolor}{rgb}{0.345, 0.345, 0.345}

\usepackage{framed}
\makeatletter
 {\par\unskip\endMakeFramed%
 \at@end@of@kframe}
\makeatother

\definecolor{shadecolor}{rgb}{.97, .97, .97}
\definecolor{messagecolor}{rgb}{0, 0, 0}
\definecolor{warningcolor}{rgb}{1, 0, 1}
\definecolor{errorcolor}{rgb}{1, 0, 0}
\newenvironment{knitrout}{}{} 

\usepackage{alltt}

\usepackage[hyphens]{url}
\usepackage{amsmath}
\usepackage{amsfonts}
\usepackage{amssymb}
\usepackage{amsthm}

\theoremstyle{plain}

\theoremstyle{definition}

\theoremstyle{remark}

\PassOptionsToPackage{hyphens}{url}\usepackage{hyperref}
\usepackage{hyperref}
\usepackage[square,numbers]{natbib}

\providecommand{\gitHash}{449717d09e492ffbf843f0acebfd46c7c2ebb6c0}

\usepackage[newitem,newenum,increaseonly]{paralist}


\usepackage[notheorems]{SharpeR}

    \usepackage{color} 
    \usepackage{enumerate} 
    \usepackage{amsmath} 
    \usepackage{amssymb} 
		\usepackage{fancyvrb} 
    \usepackage{hyperref}

    \definecolor{orange}{cmyk}{0,0.4,0.8,0.2}
    \definecolor{darkorange}{rgb}{.71,0.21,0.01}
    \definecolor{darkgreen}{rgb}{.12,.54,.11}
    \definecolor{myteal}{rgb}{.26, .44, .56}
    \definecolor{gray}{gray}{0.45}
    \definecolor{lightgray}{gray}{.95}
    \definecolor{mediumgray}{gray}{.8}
    \definecolor{inputbackground}{rgb}{.95, .95, .85}
    \definecolor{outputbackground}{rgb}{.95, .95, .95}
    \definecolor{traceback}{rgb}{1, .95, .95}
    \definecolor{red}{rgb}{.6,0,0}
    \definecolor{green}{rgb}{0,.65,0}
    \definecolor{brown}{rgb}{0.6,0.6,0}
    \definecolor{blue}{rgb}{0,.145,.698}
    \definecolor{purple}{rgb}{.698,.145,.698}
    \definecolor{cyan}{rgb}{0,.698,.698}
    \definecolor{lightgray}{gray}{0.5}
    
    \definecolor{darkgray}{gray}{0.25}
    \definecolor{lightred}{rgb}{1.0,0.39,0.28}
    \definecolor{lightgreen}{rgb}{0.48,0.99,0.0}
    \definecolor{lightblue}{rgb}{0.53,0.81,0.92}
    \definecolor{lightpurple}{rgb}{0.87,0.63,0.87}
    \definecolor{lightcyan}{rgb}{0.5,1.0,0.83}

    \definecolor{incolor}{rgb}{0.0, 0.0, 0.5}
    \definecolor{outcolor}{rgb}{0.545, 0.0, 0.0}

    \hypersetup{
      breaklinks=true,  
      colorlinks=true,
      urlcolor=blue,
      linkcolor=darkorange,
      citecolor=darkgreen,
      }
    
\providecommand{\cono}[1][\typeI]{\mathSUB{c}{#1}}

\providecommand{\dftwo}{\ssiz-\nstrat}

\IfFileExists{upquote.sty}{\usepackage{upquote}}{}
\begin{document}

\title{Inference on Achieved Signal Noise Ratio}
\author{Steven E. Pav \thanks{\email{steven@gilgamath.com}
The code to build this document is available at
\href{http://www.github.com/shabbychef/snrinf}{\normalfont\texttt{www.github.com/shabbychef/snrinf}}.
This revision was built from commit \texttt{\gitHash} of that repo.
}}

\maketitle

\begin{abstract}
We describe a procedure to perform approximate inference on the
achieved signal-noise ratio of the \txtMP under Gaussian \iid returns.
The procedure relies on a statistic similar to the Sharpe Ratio Information
Criterion. \cite{doi:10.1080/14697688.2020.1718746}
Testing indicates the procedure is somewhat conservative, but
otherwise works well for reasonable values of sample and asset universe sizes.
We adapt the procedure to deal with generalizations of the portfolio optimization problem.
\end{abstract}

\section{Introduction}

For a universe of \nstrat assets, 
we consider the portfolio optimization problem
\begin{equation}
\max_{\pportw }
\frac{\trAB{\pportw}{\pvmu}}{\sqrt{\qform{\pvsig}{\pportw}}}.
\label{eqn:opt_port_I}
\end{equation}
Here \pvmu is the expected return and \pvsig is the covariance of returns.
This problem is solved by the \txtMP, defined as 
\begin{equation}
\pportwopt \defeq \minvAB{\pvsig}{\pvmu},
\end{equation}
and any positive multiple thereof.

In practice the parameters \pvmu and \pvsig are unknown and must be estimated from the data. 
The estimation of parameters is known to deterioriate the quality of the
portfolio.  \cite{michaud1989markowitz}
The \txtSNR of the \txtMP, its mean divided by its volatility, is subject to a 
fundamental bound. \cite{pav2014qbounds,ao2017solving}
While inference on the population parameters follows from classical statistics
via the connection to Hotelling's $T^2$, little is known about performing
inference on the \txtSNR achieved by the \txtMP.
Paulsen and S\"{o}hl described the \emph{Sharpe Ratio Information Criterion} (SRIC), 
which is an approximately unbiased estimator for this quantity.  \cite{doi:10.1080/14697688.2020.1718746}
Some asymptotic confidence intervals have also been described, but these
require unreasonably large sample sizes. \cite{pav2013markowitz}
Here we fill this gap, describing confidence intervals very similar
to the SRIC and using the same approximation.
Practical construction of these bounds requires one to estimate the population
effect size.
In practice this causes the confidence intervals to be slightly conservative.


\section{The Procedure}

Assume you observe returns on \nstrat assets, 
which are independently drawn from a Gaussian distribution
$\reti[t]\sim\normlaw{\pvmu,\pvsig}$.
The population \txtMP is $\pportwopt=\minv{\pvsig}\pvmu$.
The \txtSNR of this portfolio is $\psnropt=\sqrt{\qiform{\pvsig}{\pvmu}}$.
Given \ssiz observations of returns, 
one typically estimates the population parameters via
\begin{align}
 \svmu &= \oneby{\ssiz} \sum_{1\le t \le \ssiz} \reti[t],\\
 \svsig &= \oneby{\ssiz-1} \sum_{1\le t \le \ssiz} \reti[t]\tr{\reti[t]} -
  \frac{\ssiz}{\ssiz-1}\svmu\tr{\svmu}.
\end{align}
The (sample) \txtMP is $\sportwopt=\minv{\svsig}\svmu$.
The \txtASNR of \sportwopt is defined as
\begin{equation}
  \psnr[a] \defeq \frac{\tr{\pvmu}\sportwopt}{\sqrt{\qform{\pvsig}{\sportwopt}}}.
\end{equation}
It is an unobservable random quantity that we wish to perform inference on.

The \txtSR of \sportwopt is defined as
\begin{equation}
  \ssropt \defeq \frac{\tr{\svmu}\sportwopt}{\sqrt{\qform{\svsig}{\sportwopt}}}
  = \sqrt{\qiform{\svsig}{\svmu}}.
\end{equation}
We note that $T^2=\ssiz\ssrsqopt$ is the familiar Hotelling's statistic,
which is usually prescribed to perform inference on \pvmu, but
can be used to perform inference on \psnrsqopt.
\cite{anderson2003introduction,pav_the_book}

The Sharpe Ratio Information Criterion is defined as 
\cite{doi:10.1080/14697688.2020.1718746}
\begin{equation}
  SRIC \defeq \ssropt - \frac{\nstrat-1}{\ssiz\ssropt}.
\end{equation}
Under the simplifying approximation
\begin{equation}
\label{apx:perfect_cov}
\svsig \approx \pvsig,
\end{equation}
the SRIC is unbiased for the \txtASNR:
\begin{equation}
\E{SRIC} = \E{\psnr[a]}.
\label{eqn:sric_raison_detre}
\end{equation}
Note this only holds for $\nstrat > 1$, but it is simple to express
$\E{\psnr[a]}$ when $\nstrat=1$.

Inspired by the SRIC, we seek a constant $\cono[{\typeI}]$ such that
\begin{equation}
  \Pr{\psnr[a] \le \ssropt - \frac{\cono[{\typeI}]}{\ssiz\ssropt}} = \typeI.
  \label{eqn:psnr_ci_bound}
\end{equation}
Under \apxref{perfect_cov},
\begin{equation}
  \psnr[a] 
  \approx
  \frac{\tr{\pvmu}\minv{\pvsig}\svmu}{\sqrt{\qform{\pvsig}{\wrapParens{\minv{\pvsig}\svmu}}}}
  = \frac{\tr{\pvmu}\minv{\pvsig}\svmu}{\sqrt{\qiform{\pvsig}{\svmu}}}
  = \frac{\tr{\pvmu}\minv{\pvsig}\svmu}{\ssropt}.
\end{equation}
Under the approximation we also have $\ssrsqopt\approx\qiform{\pvsig}{\svmu}$.
We note that for Gaussian returns, we can write
\begin{equation*}
  \svmu = \pvmu + \oneby{\sqrt{\ssiz}}\chol{\pvsig}\vect{z},
\end{equation*}
where $\vect{z} \sim\normlaw{\vzero,\eye}.$
Thus
\begin{align*}
  \ssrsqopt - \psnr[a]\ssropt 
  & \approx 
  \qiform{\pvsig}{\svmu} - \tr{\pvmu}\minv{\pvsig}\svmu,\\
  &=
  \tr{\wrapParens{\svmu-\pvmu}}\minv{\pvsig}\svmu,\\
  &=
  \oneby{\ssiz}
  \tr{\vect{z}}\trchol{\pvsig}
  \minv{\pvsig}
  \wrapParens{\sqrt{\ssiz}\pvmu + \chol{\pvsig}\vect{z}},\\
  &=
  \oneby{\ssiz}
  \tr{\vect{z}}
  \wrapParens{\sqrt{\ssiz}\ichol{\pvsig}\pvmu + \vect{z}},\\
  &=
  \oneby{\ssiz}
  \tr{\wrapParens{\half\sqrt{\ssiz}\ichol{\pvsig}\pvmu + \vect{z} - \half\sqrt{\ssiz}\ichol{\pvsig}\pvmu}}
     \wrapParens{\half\sqrt{\ssiz}\ichol{\pvsig}\pvmu + \vect{z} + \half\sqrt{\ssiz}\ichol{\pvsig}\pvmu},\\
  &=
  \oneby{\ssiz}\wrapParens{
    \norm{\half\sqrt{\ssiz}\ichol{\pvsig}\pvmu + \vect{z}}^2
    - \norm{\half\sqrt{\ssiz}\ichol{\pvsig}\pvmu}^2},\\
  &\sim
  \oneby{\ssiz}\wrapParens{\chisqlaw{\nstrat,\frac{\ssiz\psnrsqopt}{4}} - \frac{\ssiz\psnrsqopt}{4}}.
\end{align*}%

Now because 
$$
\psnr[a] \le \ssropt - \frac{c}{\ssiz\ssropt} 
  \Leftrightarrow c \le \ssiz \wrapParens{\ssrsqopt - \psnr[a]\ssropt},
$$
If we want this condition to hold with probability \typeI we should set
\begin{equation}
\cono[\typeI] = {\chisqqnt{1-\typeI}{\nstrat, \frac{\ssiz\psnrsqopt}{4}} - \frac{\ssiz\psnrsqopt}{4}},
\end{equation}
where \chisqqnt{q}{\df,\nctp} is the $q$ quantile of the non-central chi-square distribution
with \df degrees of freedom and non-centrality parameter \nctp.

\paragraph{Checking coverage}


Before proceeding, we check whether use of \apxref{perfect_cov}
leads to a degradation in coverage of a confidence interval implied by \ineqnref{psnr_ci_bound}.
We draw \ssiz days of returns from the $\nstrat$-variate normal distribution.
For a fixed value of \psnropt, we perform 
$1,000,000$ simulations of computing $\psnr[a]$ and \ssrsqopt, computing a
one-sided confidence bound and measuring the empirical rate of type I errors.
We then let \ssiz vary from 50 to $102,400$ days; 
we let \nstrat vary from 2 to 16;
we let \psnropt vary from $0.5\,\yrtomhalf$ to $2\,\yrtomhalf$, where we assume
252 days per year. 
We compute the lower confidence limit on $\psnr[a]$ using knowledge of the
actual \psnropt to construct \cono[\typeI].
For practical inference this would
have to be estimated, but here we are only testing conditions for which the 
approximation $\svsig\approx\pvsig$ is close enough for purposes of inference.

%

\begin{knitrout}\small
\definecolor{shadecolor}{rgb}{0.969, 0.969, 0.969}\color{fgcolor}\begin{figure}[h]
\includegraphics[width=0.975\textwidth,height=0.4875\textwidth]{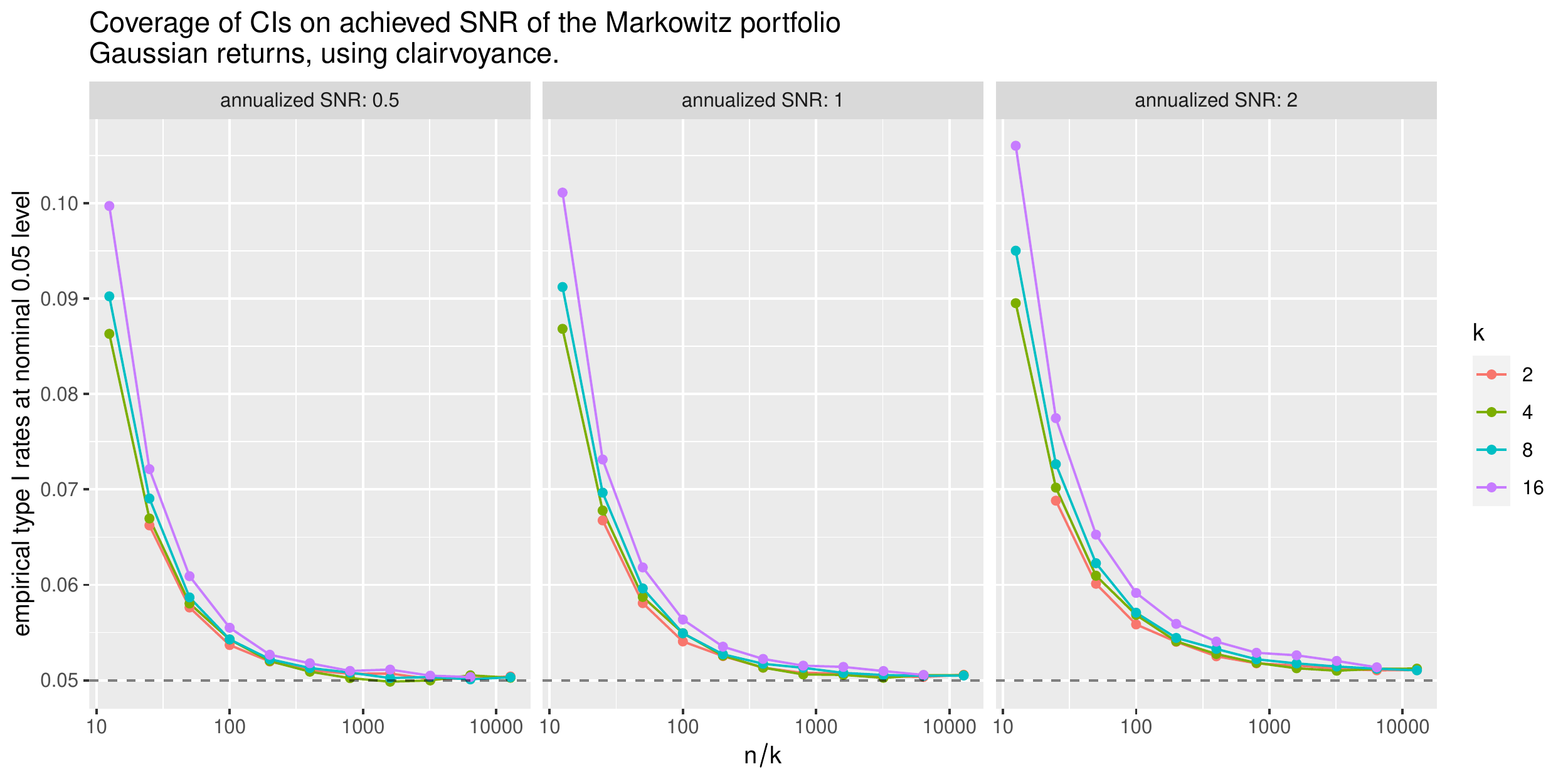} \caption[The empirical type I rate, over 1,000,000 simulations, of a one-sided confidence bound for $\psnr_{a}$ are shown for a nominal type I rate of $0.05$]{The empirical type I rate, over 1,000,000 simulations, of a one-sided confidence bound for $\psnr_{a}$ are shown for a nominal type I rate of $0.05$. The daily returns are drawn from multivariate normal distribution with varying \psnropt, \ssiz, and \nstrat. Type I rates are plotted versus $\ssiz/\nstrat$ to indicate the requisite aspect ratio to achieve near nominal coverage. }\label{fig:badci_ci_plots}
\end{figure}

\end{knitrout}
In \figref{badci_ci_plots} we plot the empirical type I rate at the nominal
0.05 level of the confidence bound. 
The main takeaway from this experiment is that the bound gives near-nominal
coverage when $\ssiz \ge 100\nstrat$ or so.


\subsection{Practical Inference}

One can construct one- or two-sided confidence intervals from \ineqnref{psnr_ci_bound} when \psnropt is known.
However, it is unknown in practice, and the constant $\cono[\typeI]$ is sufficiently sensitive to it.
To practically perform inference, there are two obvious routes: one is to
jointly perform inference on \psnropt on \psnr[a]; 
the other is to estimate \psnropt and plug it in when constructing \cono[\typeI].

For the joint estimation procedure, for some $q \in \oointerval{0}{1}$,
construct a $q \typeI$ upper bound on \psnropt.
That confidence bound can be described implicitly via the connection 
to the non-central \flaw{} distribution:
to find the one-sided confidence intervals $\ccinterval{0}{\psnr[u]}$ with
coverage $1-q\typeI$, find
\begin{equation}
\psnr[u] = \min \setwo{z}{z \ge 0,\,\,\typeI/2 \ge \ncfcdf{\wrapNeParens{\frac{\ssiz (\ssiz-\nstrat)}{\nstrat (\ssiz - 1)}}\ssrsqopt}{\nstrat,\ssiz - \nstrat}{\ssiz z^2}},
\end{equation}
where $\ncfcdf{x}{\nctdf[1],\nctdf[2]}{\ncfp}$ is the CDF of the non-central
\flaw{}-distribution with 
non-centrality parameter \ncfp and $\nctdf[1]$ and $\nctdf[2]$ degrees of freedom.
This method requires computational inversion of the CDF function. 
Then compute
$$
c = \max \wrapBraces{\left. \chisqqnt{1-\wrapNeParens{1-q}\typeI}{\nstrat,
\frac{\ssiz\psnr^2}{4}} - \frac{\ssiz\psnr^2}{4} \right|  0 \le \psnr \le \psnr[u]}.
$$
The bound $\ssropt - \frac{c}{\ssiz\ssropt}$ then should
have type I rate at most \typeI.
However, since this is a joint confidence bound the bound on \psnr[a] will be
somewhat conservative.

Another approach, which does not have guaranteed coverage, is to
estimate \psnropt from the data, and plug in that value in the computation of
\cono[\typeI].
We can perform this estimation using standard techniques, 
again via the connection of Hotelling's $T^2$ to the \flaw{} distribution.
Kubokawa, Robert and Saleh described improved methods for estimating
the non-centrality parameter given an observation of a non-central
\flaw{} statistic. \cite{kubokawa1993estimation}.
They described the following estimators for the non-centrality parameter,
which is \psnrsqopt in our case:
\begin{equation}
\label{eqn:KRS_estimators}
\begin{split}
	\delta_0 &= \frac{\wrapParens{\dftwo-2}}{\ssiz-1}\ssrsqopt - \frac{\nstrat}{\ssiz},\\
	\delta_1 &= \fmax{\delta_0,0},\\
	\delta_2 &= \fmax{\delta_0,\frac{2}{\nstrat+2}\wrapParens{\delta_0 + \frac{\nstrat}{\ssiz}}}.
\end{split}
\end{equation}
They note that $\delta_0$ is the Uniform Minimum Variance Unbiased Estimator (UMVUE) of \psnrsqopt.
However, it can be negative.
The estimators $\delta_1, \delta_2$ are non-negative, and dominate $\delta_0$ in having
lower expected squared error.
Thus the suggested procedure is to compute
$$
c = \chisqqnt{1-\typeI}{\nstrat, \frac{\ssiz\delta_2}{4}} - \frac{\ssiz\delta_2}{4},
$$
then use the bound $\ssropt - \frac{c}{\ssiz\ssropt}$.
In practice this bound seems to give slightly less conservative coverage than
the joint bound described above.
It is not clear how to find a coverage guarantee for this bound.
The quantities \ssrsqopt and \psnr[a] are not independent, 
and their asymptotic correlation is $\bigo{\ssiz^{-\halff}}$, which is only slowly shrinking. \cite{pav2013markowitz}

\paragraph{Feasible CI Coverage}

We reconsider the experiments above 
but compute feasible confidence bounds.
We use both the simultaneous CI approach with $q=0.25$; 
and plug in $\psnropt=\sqrt{\delta_2}$ to construct the bound.
In \figref{feasible_ci_plots}, we plot the empirical type I rate for both of these bounds
versus \ssiz, with facets for \psnropt.
We see that the $\delta_2$ plug-in estimator has coverage closer to the nominal $0.05$ rate.
Both bounds have issues when $\ssiz/\nstrat$ is not sufficiently large, a problem stemming
from the poor quality of the approximation $\svsig\approx\svsig$, and which was seen above.
However, here we see closer to nominal coverage for larger \nstrat for both methods.
It is not clear how the coverage will behave for larger $\ssiz/\nstrat$, though
that seems like an unlikely problem in practice.
%
\begin{knitrout}\small
\definecolor{shadecolor}{rgb}{0.969, 0.969, 0.969}\color{fgcolor}\begin{figure}[h]
\includegraphics[width=0.975\textwidth,height=0.4875\textwidth]{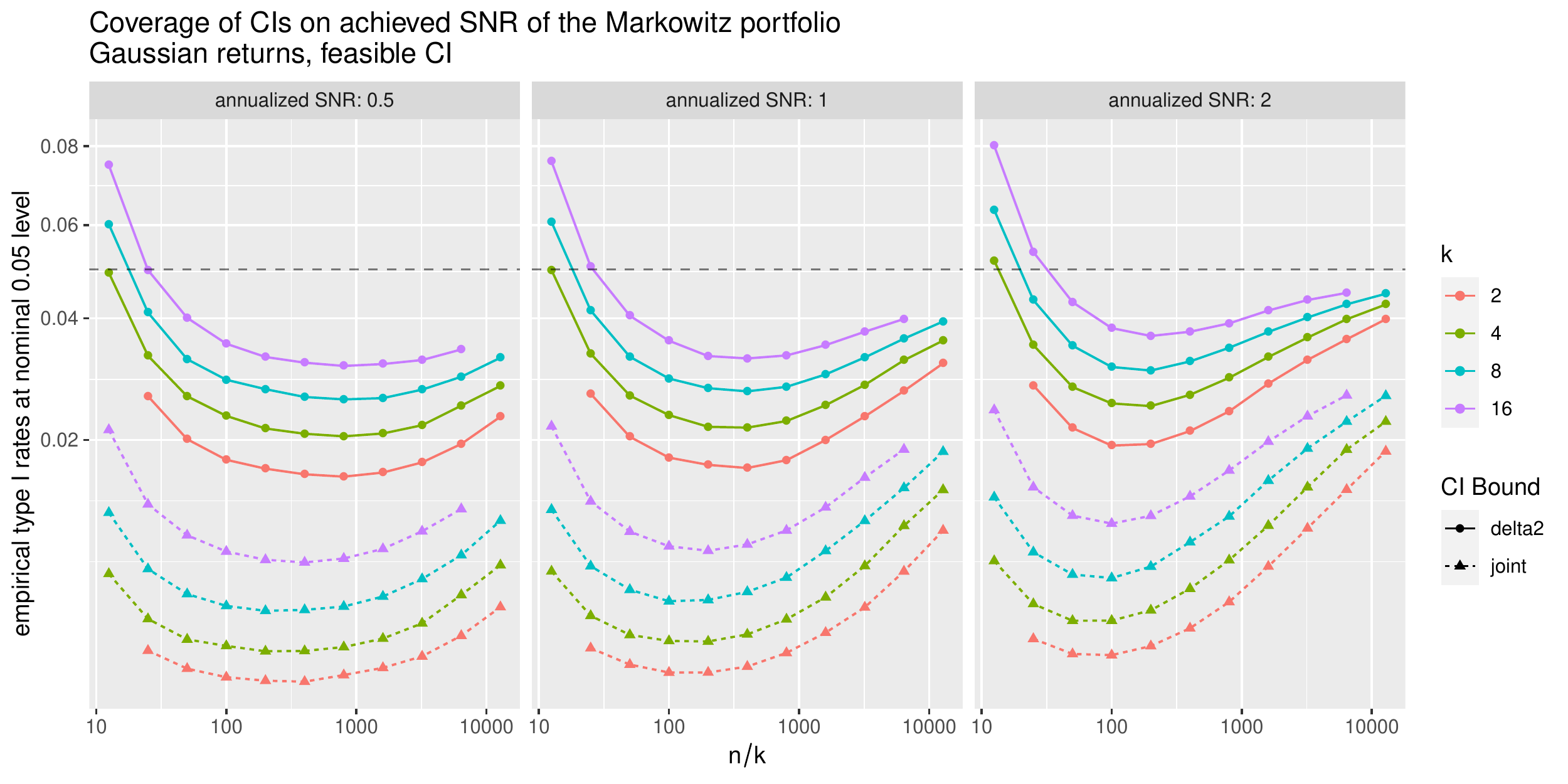} \caption[The empirical type I rate, over 1,000,000 simulations, of two feasible one-sided confidence bounds for $\psnr_{a}$ are shown for a nominal type I rate of $0.05$]{The empirical type I rate, over 1,000,000 simulations, of two feasible one-sided confidence bounds for $\psnr_{a}$ are shown for a nominal type I rate of $0.05$. The daily returns are drawn from multivariate normal distribution with varying \psnropt, \ssiz, and \nlatf. The $y$ axis is drawn in square root scale to show detail. }\label{fig:feasible_ci_plots}
\end{figure}

\end{knitrout}

\subsection{Hedged Portfolios}

Now we generalize the portfolio problem of \eqnref{opt_port_I} to add a hedging constraint.
So consider the constrained portfolio optimization problem on \nstrat assets,
\begin{equation}
  \max_{\substack{\hejG\pvsig \pportw = \vzero,\\ \qform{\pvsig}{\pportw} \le \Rbuj^2}}
\frac{\trAB{\pportw}{\pvmu} - \rfr}{\sqrt{\qform{\pvsig}{\pportw}}},
\label{eqn:opt_port_hedge}
\end{equation}
where $\hejG$ is an $\nstrathej \times \nstrat$ matrix of rank \nstrathej, and,
as previously, \pvmu, \pvsig are the mean vector and covariance matrix, 
\rfr is the risk-free rate, and $\Rbuj > 0$ is a risk `budget'. 
We can interpret
the \hejG constraint as stating that the covariance of the returns of
a feasible portfolio with the returns of a portfolio whose weights are in
a given row of \hejG shall equal zero. 
In the garden variety application of this problem, \hejG consists of 
\nstrathej rows of the identity matrix; 
in this case, feasible portfolios are \emph{hedged} with respect 
to the \nstrathej assets selected by \hejG
(although they may hold some position in the hedged assets).
We use ``hedged'' to mean a portfolio with zero covariance 
against some other portfolio(s).

The
solution to this problem, via the Lagrange multiplier technique,
is
\begin{equation*}
\pportwoptHej{\hejG}
= c \wrapParens{\minv{\pvsig}{\pvmu} -
  \qiform{\wrapParens{\qoform{\pvsig}{\hejG}}}{\hejG}\pvmu}.
\end{equation*}
When $\rfr > 0$, the unique solution is found by setting $c$ so that the risk budget is an equality.
Note that, up to scaling, $\minv{\pvsig}\pvmu$ is the unconstrained optimal
portfolio, and thus the imposition of the \hejG constraint only changes
the unconstrained portfolio in assets corresponding to columns of \hejG 
containing non-zero elements. In the garden variety application where
\hejG is a single row of the identity matrix, the imposition of the
constraint only changes the holdings in the asset to be hedged (modulo
changes in the leading constant to satisfy the risk budget).


The squared \txtSNR of the optimal portfolio we write as 
\begin{equation}
\Hejpsnrsqopt{\hejG} 
\defeq 
\qiform{\pvsig}{\pvmu} - \qiform{\wrapParens{\qoform{\pvsig}{\hejG}}}{\wrapParens{\hejG\pvmu}}.
\label{eqn:psnr_Gcons}
\end{equation}
The sample optimal portfolio is given by 
\begin{equation*}
  \sportwoptHej{\hejG}
= c \wrapParens{\minv{\svsig}{\svmu} -
  \qiform{\wrapParens{\qoform{\svsig}{\hejG}}}{\hejG}\svmu}.
\end{equation*}
The squared \txtSR of this portfolio is
\begin{equation}
\Hejssrsqopt{\hejG}
= \qiform{\svsig}{\svmu} -
  \qiform{\wrapParens{\qoform{\svsig}{\hejG}}}{\wrapParens{\hejG\svmu}}.
\label{eqn:ssr_Gcons} 
\end{equation}
The \txtASNR of this portfolio is
\begin{equation}
  \psnr[a] = 
  \frac{\tr{\pvmu}\sportwoptHej{\hejG}}{\sqrt{\qform{\pvsig}{\sportwoptHej{\hejG}}}}.
\end{equation}

Define:
\begin{align*}
\ssrsqoptG{\hejG} &\defeq \qiform{\wrapParens{\qoform{\svsig}{\hejG}}}{\wrapParens{\hejG\svmu}}.
\end{align*}
Giri showed that conditional on observing \ssrsqoptG{\hejG},
{\small
\begin{equation}
\acondb{
	\frac{\ssiz}{\ssiz-1}\frac{\ssiz - \nstrat}{\nstrat - \nstrathej}\frac{\Hejssrsqopt{\hejG}}{1 +
  \frac{\ssiz}{\ssiz-1}\ssrsqoptG{\hejG}} 
}{\ssrsqoptG{\hejG}}
	\sim
\ncflaw{\nstrat-\nstrathej,\ssiz-\nstrat,\frac{\ssiz}{1 +
\frac{\ssiz}{\ssiz-1}\ssrsqoptG{\hejG}} 
	\Hejpsnrsqopt{\hejG}
	},
\label{eqn:giri_done_I}
\end{equation}%
}%
where \ncflaw{\df[1],\df[2],\ncfp} is the non-central \flaw{}-distribution
with \df[1], \df[2] degrees of freedom and non-centrality parameter
\ncfp.  \cite{giri1964likelihood,pav_the_book}


Now we apply \apxref{perfect_cov}, and complete the square as we did in the unhedged case, to find that
{\scriptsize
\begin{align*}
  \Hejpsnrsqopt{\hejG} - \psnr[a]\sqrt{\Hejpsnrsqopt{\hejG}} 
  & \approx 
  \tr{\wrapParens{\svmu-\pvmu}}\minv{\pvsig}\svmu -
  \tr{\wrapParens{\svmu-\pvmu}}\tr{\hejG}
  \minv{\wrapParens{\qoform{\pvsig}{\hejG}}}\hejG\svmu,\\
  &=
  \oneby{\ssiz}
  \wrapBracks{
    \qform{\wrapParens{\eye - \trchol{\pvsig}\tr{\hejG}\minv{\wrapParens{\qoform{\pvsig}{\hejG}}}\hejG\chol{\pvsig}}}{%
      \wrapParens{\half\sqrt{\ssiz}\ichol{\pvsig}\pvmu + \vect{z}}}}\\
    &\phantom{=}\,
  -\oneby{4\ssiz}
  \wrapBracks{
    \qform{\wrapParens{\eye - \trchol{\pvsig}\tr{\hejG}\minv{\wrapParens{\qoform{\pvsig}{\hejG}}}\hejG\chol{\pvsig}}}{%
      \wrapParens{\sqrt{\ssiz}\ichol{\pvsig}\pvmu}}},
\end{align*}%
}%
where $\vect{z} \sim\normlaw{\vzero,\eye}.$
Now note that the matrix 
$$\Mtx{A} = {\eye - \trchol{\pvsig}\tr{\hejG}\minv{\wrapParens{\qoform{\pvsig}{\hejG}}}\hejG\chol{\pvsig}}$$
is idempotent with rank $\nstrat - \nstrathej$.
Thus a quadratic form in \Mtx{A} follows a non-central $\chi^2$ distribution\footnote{\emph{n.b.} the standard definition of
non-centrality parameter in the time Graybill and Marsaglia wrote their paper is different from the one we use today by a factor of
$1/2$.} with degrees of freedom equal to the rank of \Mtx{A}. \cite[Theorem 2]{10.2307/2237227}
Thus
\begin{equation*}
\Hejpsnrsqopt{\hejG} - \psnr[a]\sqrt{\Hejpsnrsqopt{\hejG}} \sim
  \oneby{\ssiz}\wrapParens{\chisqlaw{\nstrat - \nstrathej,\frac{\ssiz\Hejpsnrsqopt{\hejG}}{4}} - \frac{\ssiz\Hejpsnrsqopt{\hejG}}{4}}.
\end{equation*}

Then, as in the unhedged case, we have
\begin{equation}
  \Pr{\psnr[a] \le \sqrt{\Hejssrsqopt{\hejG}} - \frac{\cono[{\typeI}]}{\ssiz\sqrt{\Hejssrsqopt{\hejG}}}} = \typeI,
\end{equation}
if we let
$$
\cono[\typeI] = {\chisqqnt{1-\typeI}{\nstrat - \nstrathej, \frac{\ssiz\Hejpsnrsqopt{\hejG}}{4}} - \frac{\ssiz\Hejpsnrsqopt{\hejG}}{4}}.
$$
To perform feasible inference one will need to estimate \Hejpsnrsqopt{\hejG}. 
Again this will be via the connection to a non-central \flaw{}-distribution, \eqnref{giri_done_I}.
One can either find an upper quantile directly, or use a KRS-type estimator, which 
for the hedged case are 
\begin{equation}
\label{eqn:hedged_KRS_estimators}
\begin{split}
  \delta_0 &= \frac{\wrapParens{\dftwo-2}}{\ssiz-1}\Hejssrsqopt{\hejG} - \frac{\nstrat-\nstrathej}{\ssiz}\wrapParens{1 +
  \frac{\ssiz}{\ssiz-1}\ssrsqoptG{\hejG}},\\
	\delta_1 &= \fmax{\delta_0,0},\\
  \delta_2 &= \fmax{\delta_0,\frac{2}{\nstrat-\nstrathej+2}\frac{\dftwo-2}{\ssiz-1}\Hejssrsqopt{\hejG}}.
\end{split}
\end{equation}

\paragraph{Checking coverage}


As in the unhedged case, we first perform simulations where the population parameter \Hejpsnrsqopt{\hejG} is known,
to assess the effects of \apxref{perfect_cov}.
In our simulations, we set $\nstrathej = \nstrat/2$, and let $\hejG$ be the first \nstrathej rows of the identity matrix.
We set $\pvmu=c \vone$ and $\pvsig=\eye$.
We perform $100,000$ simulations for different values of $\Hejpsnrsqopt{\hejG}$, $\nstrat$ and $\ssiz$,
computing \psnr[a] for the hedged portfolio, as well as \Hejssrsqopt{\hejG} and \ssrsqoptG{\hejG}.
We compute the lower 0.05 bound using knowledge of \Hejpsnrsqopt{\hejG} and compute the empirical type I rate
over the $100,000$ simulations, which we plot versus $\ssiz/\nstrat$ in \figref{hejci_plots}.
Again we see that the nominal type I rate is nearly achieved when $\ssiz > 100\nstrat$ or so.

\begin{knitrout}\small
\definecolor{shadecolor}{rgb}{0.969, 0.969, 0.969}\color{fgcolor}\begin{figure}[h]
\includegraphics[width=0.975\textwidth,height=0.4875\textwidth]{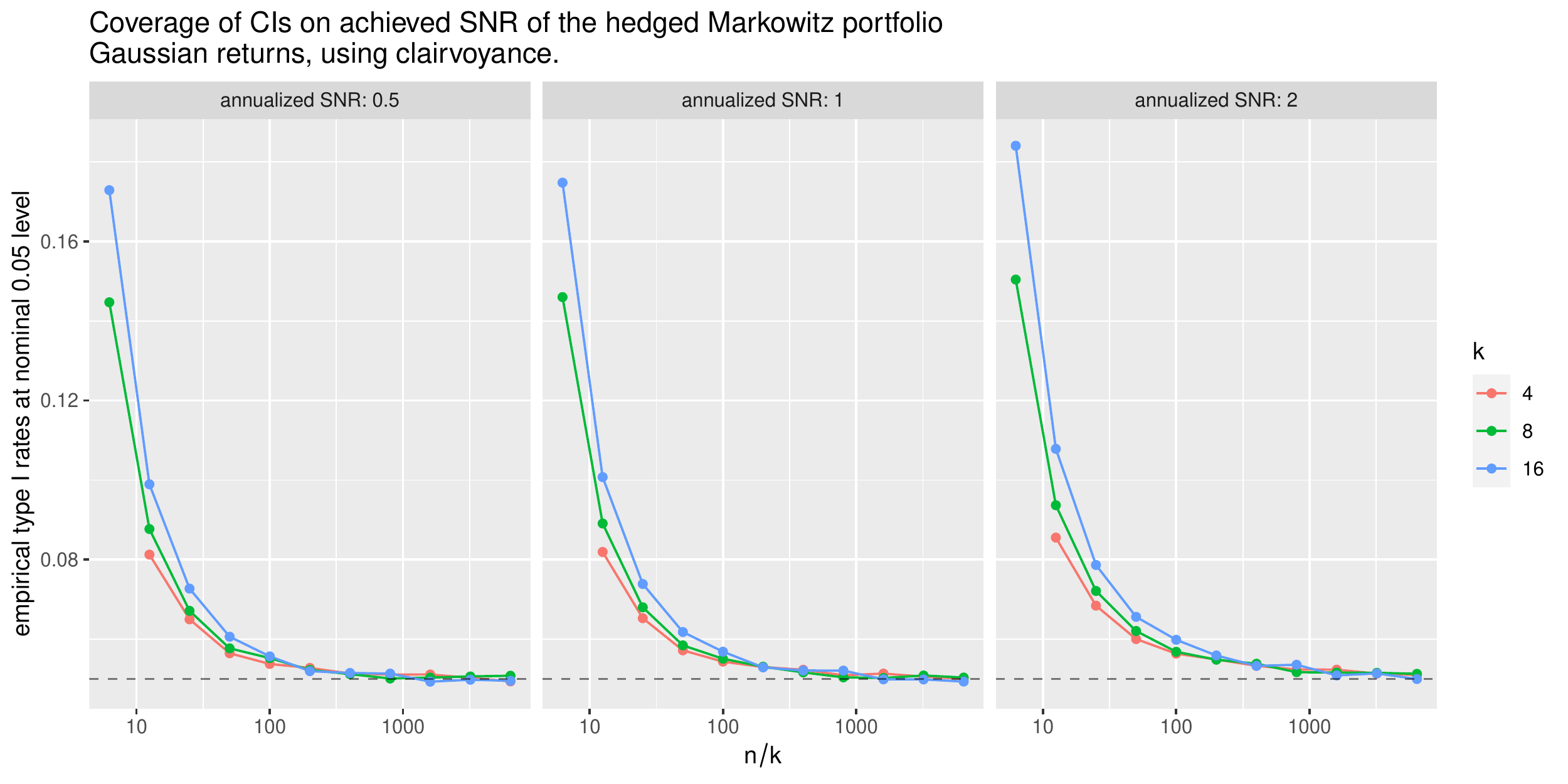} \caption{The empirical type I rate, over 100,000 simulations, of a one-sided confidence bound for $\psnr_{a}$ are shown for the hedged portfolio problem.  We set $\nstrathej=\nstrat/2$ and take $\pvmu\propto\vone$. The SNR in the facet titles refers to $\sqrt{\Hejpsnrsqopt{\hejG}}$. }\label{fig:hejci_plots}
\end{figure}

\end{knitrout}

As above we analyze the data from the hedged experiments, but compute feasible confidence bounds.
We use both the simultaneous CI approach with $q=0.25$; 
and plug in $\Hejpsnrsqopt{\hejG}=\sqrt{\delta_2}$ to construct the bound.
In \figref{feasible_hejci_plots}, we plot the empirical type I rate for both of these bounds
versus \ssiz, with facets for \psnropt and \nstrat.
Once again, the $\delta_2$ plug-in estimator has coverage closer to the nominal $0.05$ rate,
and both bounds are anti-conservative when $\ssiz/\nstrat$ is not sufficiently large.
\begin{knitrout}\small
\definecolor{shadecolor}{rgb}{0.969, 0.969, 0.969}\color{fgcolor}\begin{figure}[h]
\includegraphics[width=0.975\textwidth,height=0.4875\textwidth]{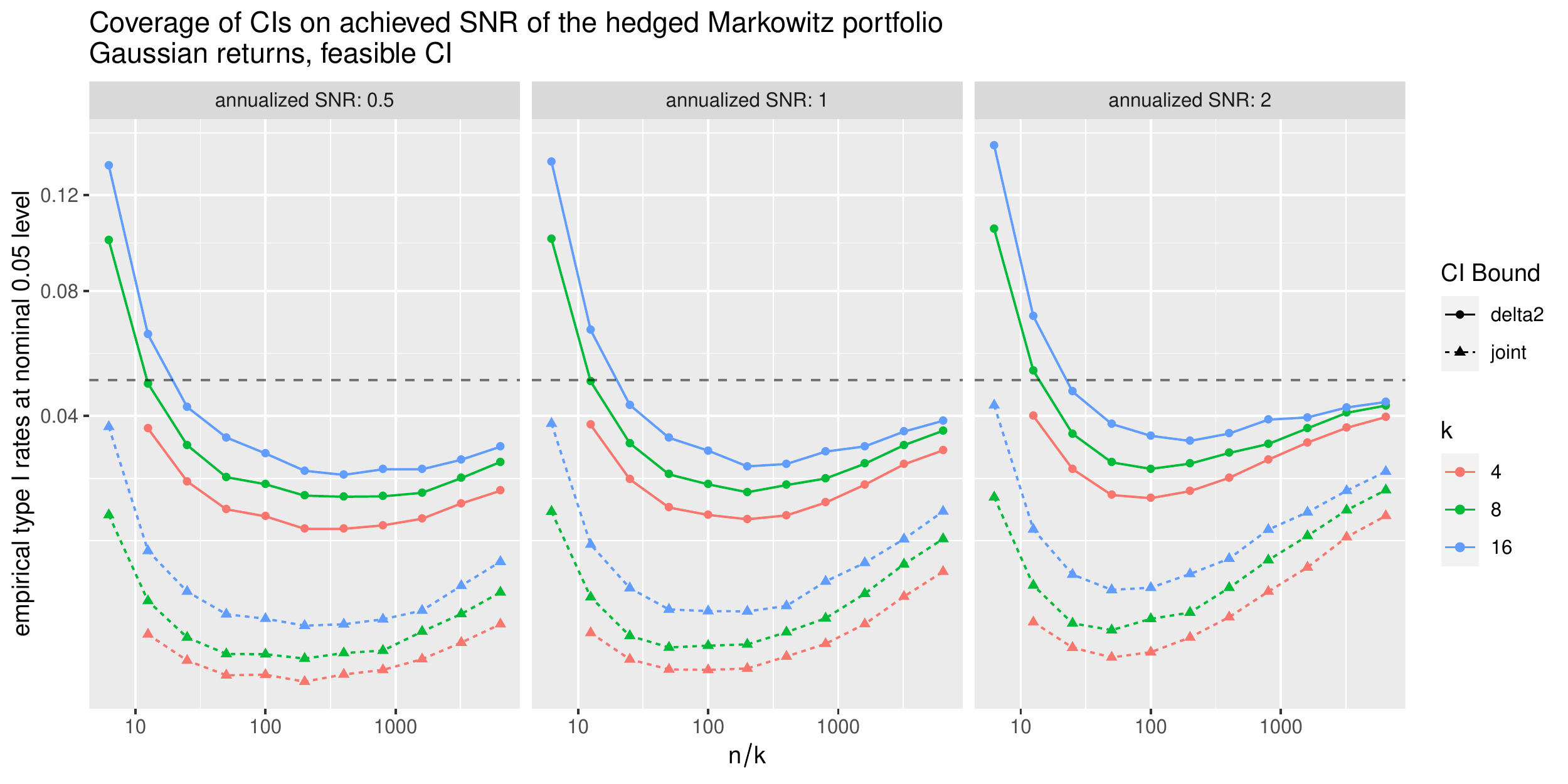} \caption[The empirical type I rate, over 100,000 simulations, of two feasible one-sided confidence bounds for $\psnr_{a}$ are shown for a nominal type I rate of $0.05$]{The empirical type I rate, over 100,000 simulations, of two feasible one-sided confidence bounds for $\psnr_{a}$ are shown for a nominal type I rate of $0.05$. The daily returns are drawn from multivariate normal distribution with varying \psnropt, \ssiz, and \nlatf. The $y$ axis is drawn in square root scale to show detail. }\label{fig:feasible_hejci_plots}
\end{figure}

\end{knitrout}



\section{Examples}

\paragraph{Fama French 4 Factor Returns}
\label{example:3400185f-3864-4011-94df-5f756d81c044}

We consider a portfolio constructed on the `Market', size (SMB), value (HML)
and momentum (HMD) portfolios described by Fama and French, \emph{inter alia},
with data compiled and published by Kenneth French.  \cite{Fama_French_1992,Carhart_1997,french_data_library}
The set consists of $\ssiz=1104\,\moto{}$ of data, from 
1927 through 2018.917. 
We observe $\ssrsqopt=0.098\,\moto{-1}$.
From this we compute $\delta_2 = 0.094\,\moto{-1}$.
Plugging in $\sqrt{\delta_2}$ for \psnropt 
we compute a two-sided $95\%$ confidence bound on \psnr[a] as 
$\ccinterval{0.234}{0.353}\,\moto{-\halff}$.
By comparison, via the connection to the \flaw{} distribution,
we compute $95\%$ confidence intervals on \psnropt as
$\ccinterval{0.247}{0.369}\,\moto{-\halff}$.

Next we consider the imposition of a constraint that the portfolio should be ``hedged against the Market''.
This corresponds to $\nstrathej=1$ and \hejG is the row of the identity matrix corresponding to the Market factor.
We compute
$$
\ssrsqopt=0.098\,\moto{-1},\quad
\ssrsqoptG{\hejG}=0.03\,\moto{-1},\quad
\Hejssrsqopt{\hejG}=0.068\,\moto{-1}.
$$
With $\ssiz=1104, \nstrat=4$, we 
compute the three KRS estimators of \Hejpsnrsqopt{\hejG}, which all take the same value,
$\delta_0=\delta_1=\delta_2=0.065\moto{-1}.$
From this we compute a one-sided $95\%$ confidence bound on 
\psnr[a]
to be 
$\cointerval{0.195}{\infty}\,\moto{-\halff}$.


\section{Discussion}

Testing indicates the confidence bound exhibits closer to nominal
coverage than the known asymptotic bounds for reasonable \ssiz and \nstrat.
Further work should naturally focus on mitigating the effects of the
approximation $\svsig\approx\pvsig$, and finding a coverage guarantee
of the plug-in estimator. 
We also anticipate that this confidence bound procedure can be adapted
to deal with 
conditional expectation models.


\bibliographystyle{plainnat}
\bibliography{snrinf}

\end{document}